\documentclass[12pt]{iopart}

\usepackage{latexsym}
\usepackage[T1]{fontenc}
\usepackage[utf8x]{inputenc}
\usepackage{latexsym}
\usepackage{times}
\usepackage{graphicx,amsfonts,amsbsy,amssymb,amsthm}
\usepackage{amsopn}
\usepackage{color}
\usepackage{cite}
\usepackage{hyperref}

\newcommand{\E}{\mathbf{E}}

\newcommand{\blambda}{\boldsymbol{\lambda}}
\newcommand{\bt}{{\boldsymbol t}}

\def\={\;=\;} \def\+{\,+\,}

\def\mc {\mathcal}

\newtheorem*{prop*}{Proposition}

\begin{document}

\title[The probability distribution of spectral moments for the G$\beta$E]{The probability distribution of spectral moments for the Gaussian $\beta$-Ensembles}
\author{Tomasz Maci\k{a}\.{z}ek$^{1}$, Christopher H. Joyner$^{2}$ and Uzy Smilansky$^{3}$}

\address{$^{1}$Center for Theoretical Physics, Polish Academy of Sciences, Al. Lotnik\'ow 32/46, 02-668 Warsaw, Poland.}
\address{$^{2}$School of Mathematical Sciences, Queen Mary University of London, London, E1 4NS, UK.}
\address{$^{3}$Department of Physics of Complex Systems, Weizmann Institute of Science, Rehovot 7610001, Israel.}
\ead{\mailto{maciazek@cft.edu.pl}, \mailto{c.joyner@qmul.ac.uk}, \mailto{uzy.smilansky@weizmann.ac.il}
}
\begin{abstract}
We show how to obtain the joint probability distribution of the first two spectral moments for the G$\beta$E random matrix ensembles of any matrix dimension $N$. This is achieved via a simple method which utilises two complementary invariants of the domain of the spectral moments. Our approach is significantly different from those employed previously to answer related questions and potentially offers new insights. We also discuss the problems faced when attempting to include higher spectral moments.
\end{abstract}
\maketitle

\section {Introduction}
\label{introduction}
In the following we consider the \emph{spectral moments}, which are given (up to normalisation) by the \emph{traces} of an $N \times N$ hermitian matrix $M$ with spectrum $\blambda = (\lambda_1,\ldots, \lambda_N)$
\begin{equation}
\label{eq:themap}
t_r= {\tr M}^r =\sum_{k=1}^N \lambda_k^r. 
\end{equation}
Terms \emph{traces} and \emph{spectral moments} appear frequently in the literature and we shall use them interchangeably here. In particular we shall focus on the first $N$ traces, denoting the vector $\bt = (t_1,\ldots,t_N)$.

The spectral moments play an important in role random matrix theory, where they are typically used to infer global properties of the eigenvalues. Perhaps the most famous application is Wigner's semicircle law, which can be deduced from the ensemble averaged spectral moments $\E[t_n]$ for so-called \emph{Wigner matrices} (see e.g. \cite{Anderson-book}) in the limit of large matrix size \cite{Wigner-1955,Wigner-1958}. They are also crucial quantities in Mathematical Physics as they characterise the transport properties of quantum chaotic systems (see for example \cite{Mezzadri1,Mezzadri2} and references therein).

In this article we shall be concerned with the Gaussian $\beta$ Ensemble (G$\beta$E) of size $N$. By definition, the Gaussian ensembles are the ensembles of self adjoint matrices of size $N \times N$, whose entries are of the form
$M_{i j}=\sum_{\alpha=0}^{\beta-1}M_{ij;\alpha}e_{\alpha}$. The coefficients $M_{ij;\alpha}$ being
real parameters and $e_{\alpha}$ are the units of the three potential algebras: real ($\beta = 1)$,
complex ($\beta=2$) and real-quaternion ($\beta=4$), satisfying $e_0^2 =1$ and $e_\alpha^2 = -1 \ \forall\  \alpha>0$.
Choosing the real coefficients $M_{ij;\alpha}$ independently from a Gaussian distribution with zero mean and
variance $\E(M_{ij;\alpha}^2) = (1+\delta_{ij})/(2\beta)$ we obtain the Gaussian orthogonal, unitary and
symplectic ensembles (GOE, GUE and GSE) for $\beta=1,2$ and 4 respectively (see e.g. Chapter 3 of \cite{mehtabook}). Here, the joint probability distribution function (JPDF) of the spectrum is given by
\begin{equation}
\label{spectrJPD1}
P^{(N)}_{\beta} (\blambda)=  C^{(N)}_{\beta} \prod_{\mu<\nu}|\lambda_{\mu}-\lambda_{\nu}|^{\beta}
{\rm exp}\left(-\frac{\beta}{2} \sum_{\mu}\lambda_{\mu}^2\right ),
\end{equation}
where $C^{(N)}_{\beta}$ is a normalisation constant. Allowing variable $\beta$ to take arbitrary real positive values, we obtain the definition of G$\beta$E JPDF. There also exists a tridiagonal matrix model for arbitrary $\beta$ introduced by Dumitriu and Edelman \cite{Dumitriu}. Beyond the  semicircle law, it was first shown by Johansson \cite{Johansson} (and later by Dumitriu and Edelman using their tridiagonal model \cite{Dumitriu-2006}) that the fluctuations of the traces around their mean are Gaussian distributed with a covariance structure given in terms of the Chebyshev polynomials. Moreover, the variance is less than one would expect from normal central limit theorem arguments due to to the rigidity of the eigenvalues, coming from the Vandermonde factor in (\ref{spectrJPD1}). Large deviation estimates for these fluctuations have also been obtained by Cabanal-Duvillard and Guionnet \cite{Cabanal-Duvillard-2001} using Dyson Brownian motion.

Similar results showing Gaussian fluctuations have also been attained for more general Wigner matrices \cite{Sinai,Khorunzhy-1996,Bai-2005}. Moreover, the covariance structure identified by Johansson has also been shown to appear for matrices with correlated entries by Schenker and Schulz-Baldes \cite{Schenker-2007}. In particular \cite{Schenker-2007} provides a good overview of the above results. In addition we feel it is worthwhile mentioning the well-known analogous results of Diaconis and Shahshahani \cite{Diaconis-1994}, which describe the Gaussian nature of fluctuations of the traces for the CUE\footnote{The Circular Unitary Ensemble, which is the ensemble of self-adjoint unitary matrices with complex entries.}. Similar variance estimates were also attained by Haake et. al. \cite{Haake-1996} for the canonical circular ensembles and very recently Webb has gone further, extending the Diaconis and Shahshahani results to the C$\beta$E \cite{Webb-2015}.

The analysis of global fluctuations for random matrices is typically performed in the limit of large $N$. Not surprisingly, much less is known about the intricacies of the spectral moments for finite matrix sizes, however there has been progress in this direction. For instance, Ledoux \cite{Ledoux-2009} showed there exists a recursion relation for the average moments $\E[t_n]$ of the canonical Gaussian ensembles, extending the previous work of Harer and Zagier for the GUE \cite{Harer-1986}. From this one may in principle calculate $\E[t_n]$ up to all orders in $N$. For general $\beta$ Witte and Forrester have recently succeeded in deriving expressions for $\E[t_n]$ up to $\mathcal{O}(N^{-6})$ using loop equations \cite{Witte-2014}.

In the following we derive an exact expression for the JPDF of the first two traces $P(t_1,t_2)$ for any $N$. Note that the second spectral moment is of particular importance as it is equivalent to the Hilbert-Schmidt norm for self-adjoint matrices. We would like to emphasise, that in contrast to previous methods which either utilise the properties of the eigenvalues or the matrix elements directly, we work entirely within the space of traces. The transformation to this framework is presented in Section \ref{JPDF for the traces} in which we show how the JPDF $Q(\bt)$ for the first $N$ traces arises. The domain for which $Q(\bt)$ is valid is analysed and shown to obey two important symmetries. These directly lead to our main result, presented in Section \ref{JPDF for t1t2}, which is the JPDF for $t_1$ and $t_2$. From this we calculate marginal distributions and covariances in Section \ref{Marginal distributions} and Section \ref{sec:correlations} respectively. Of course, one would like to extend the approach to higher traces, however the main difficulty with this approach lies within the complexity of the integration domain. This obstruction is detailed in Section \ref{Sec: Higher spec moments} and we summarise our findings in Section \ref{Conclusions}.


\section{The joint probability distribution for the trace variables}\label{JPDF for the traces}

The starting point for our discussion is the spectral joint probability distribution of the eigenvalues for the G$\beta$E (\ref{spectrJPD1}). A natural approach for finding the distribution of traces is to map the space of eigenvalues $\blambda=(\lambda_1,\dots,\lambda_N)$ to the space of traces $\bt=(t_1,\dots,t_N)$ using (\ref{eq:themap}), and to restrict the map to the region in the $\bt$ space where it is injective. Note that every $t_k$ is a symmetric polynomial in $(\lambda_1,\dots,\lambda_N)$, hence the image of the map is invariant under permutations of the eigenvalues. This allows us to restrict the domain of the map to a   single sector in the $\blambda$ space where the eigenvalues are ordered, for example $\lambda_1<\lambda_2<\dots<\lambda_N$. The image of the map is given by the region  where the Jacobian is positive, i.e. the boundary of the image is a connected component of the zero set of the Jacobian as a function of $\bt$
\[
\left|\frac{\partial \bt}{\partial\blambda}\right|(\bt)=0.
\]
The indicator function corresponding to the region within the boundary will be denoted by $\chi(\bt)$.

Expressing the Jacobian in terms of the traces is a relatively easy task, since from (\ref{eq:themap}) the elements of the transformation are given by $\frac{\partial t_k}{\partial \lambda_m} = k \lambda_m^{k-1}$ and therefore
\begin{equation}\label{jacobian}
\left|\frac{\partial \bt}{\partial\blambda}\right|=N!\det(V(\blambda)).
\end{equation}
Here $V$ is the familiar Vandermonde matrix
\begin{equation}\label{Vandermonde}
V(\blambda)=
\left(\begin{array}{cccc}
1 & 1 & \cdots & 1 \\
\lambda_1 & \lambda_2 & \cdots & \lambda_N \\
\vdots & \vdots & \ddots & \vdots \\
\lambda_1^{N-1} & \lambda_2^{N-1} & \cdots & \lambda_N^{N-1} \end{array}\right)
\end{equation}
and so $\det(V) = \prod_{\mu < \nu} | \lambda_{\mu} - \lambda_{\nu}|$, which appears in (\ref {spectrJPD1}). Furthermore, $G(\bt) = \det(V) = \sqrt{\det(VV^{\intercal})}$, with
\[
VV^{\intercal} =
\left(\begin{array}{cccc}
t_0 & t_1 & \cdots & t_{N-1} \\
t_1 & t_2 & \cdots & t_N \\
\vdots & \vdots & \ddots & \vdots \\
t_{N-1} & t_N & \cdots & t_{2N-2} \end{array}\right)
\]
and $t_0=N$  (see \cite{Dunne-1993,Vivo-2008} for examples where this identity is used in other contexts). At this point $G(\bt)$ is expressed entirely in terms of the traces, as desired, however it includes
traces of higher degree than $N$, which are themselves functions of the traces $t_n, \ 1\le n\le N$. The expressions for $t_{N+r}$ in terms of the first $N$ $t_n$, whilst complicated, can be written down explicitly. They originate from the characteristic polynomial
$\Phi(x) := \det(xI - M) = \sum_{k=0}^N c_k x^{N-k}$, with $c_0=1$. For any eigenvalue $\lambda_\nu$ we
have $\Phi(\lambda_\nu)=0$ and thus it follows that
\begin{equation}
\label{poleq}
t_{N+r}= \ -\left [\sum_{k=1}^{N} c_k t_{N+r-k}\right ].
\end{equation}
Newton's identities express the coefficients $c_k$ in terms of the $t_n, n\le k$ via the determinant
\begin{equation}
\label{newteq}
c_k =\frac{(-1)^k}{k!}
\left|\begin{array}{ccccc}
t_1 & 1   & 0 & \cdots & 0 \\
t_2 & t_1 & 2 & \cdots & 0 \\
\vdots & \vdots & \ddots &  \ddots & \vdots\\
t_{k-1} &  t_{k-2} & \cdots  & t_1& k-1\\
t_{k} & t_{k-1} & \cdots & t_2 & t_{1} \end{array}\right|.
\end{equation}
Therefore, using a combination of the relations (\ref {poleq}) and (\ref{newteq}) one may write $G(\bt)$ explicitly in terms of the $N$ traces which are the components of  $\bt$. Clearly $\Delta = G(\bt)^2$ is nothing but the discriminant of $\Phi(x)$ expressed as a function of $\bt$. For this reason, we shall refer to the domain supported by $\chi(\bt)$ as the {\it discriminant domain}.

Computing now the expectation of any function $f(t_1,\dots,t_n)$ using (\ref{spectrJPD1}), and remembering that there are $N!$ ordered sectors in the $\blambda$ space, we have
\begin{eqnarray*}
\E[{f}]=N!C^{(N)}_{\beta}\int_{\lambda_1<\dots<\lambda_N}d\blambda f(\bt(\blambda)) \prod_{\mu<\nu}(\lambda_{\mu}-\lambda_{\nu})^{\beta} {\rm exp}\left(-\frac{\beta}{2} \sum_{\mu}\lambda_{\mu}^2\right ) = \\ =N!C^{(N)}_{\beta}\int d\bt \left|\frac{\partial \blambda}{\partial\bt}\right|\chi(\bt)f(\bt)G(\bt)^\beta\exp\left(-\frac{\beta}{2} t_2\right ).
\end{eqnarray*}
Using expression (\ref{jacobian}) for the Jacobian leads us finally to the JPDF for the traces
\begin{equation}
\label{traceJPD}
Q^{(N)}_{\beta} (\bt) =  C^{(N)}_{\beta}\chi(\bt)G(\bt)^{\beta-1}\exp\left(-\frac{\beta}{2} t_2\right).
\end{equation}
We mention in passing here the above form was shown by two of the authors to obey a suitable Fokker-Planck equation, leading to previously unnoticed identities involving the derivatives of traces larger than $N$ \cite{Joyner-2015}. Also, an equivalent formula in terms of the secular coefficients $c_k$ was calculated by Bakry and Zani \cite{Bakry-2014}.

The form of (\ref{traceJPD}) seems especially attractive for the GOE ($\beta = 1$) where the Vandermonde determinant disappears. Although, of more significance is the fact that $\chi(\bt)$ is independent of the ensemble under consideration - one may consider
matrices with non-Gaussian elements, or even correlated elements, and the function $\chi(\bt)$ will remain the same. Unfortunately, this discriminant domain is not at all simple (see e.g. \cite {Meymann} for an review of its structure). To get a flavour of this difficulty, consider the following Cauchy-Schwarz inequalities for the traces $t_r$ with $r=p+q$,
\begin{equation}
t_r =\sum_{k=1}^N \lambda_k^p\lambda_k^q \le \left (\sum_{k=1}^N \lambda_k^{2p}\sum_{k=1}^N \lambda_k^{2q}\right )^{\frac{1}{2}} = \left (t_{2p}t_{2q}\right )^{\frac{1}{2}}\ .
\end{equation}
In particular, taking $p=r$ and $q=0$ we get
\begin{equation}\label{schwartz_even}
t_r^2 \le N t_{2r},
\end{equation}
since $t_0=N$. These inequalities are optimal, since they are satisfied as equalities when the spectrum is entirely degenerate. However they are not sufficient to define the discriminant domain sharply enough, and to turn the integration (\ref {traceJPD}) transparent or practical. Even using additional more complicated inequalities does not improve the situation. For example, one may take $\sum_{k,m}(\lambda_k-\lambda_m)^{2 n} \ge 0$, then expanding the binomials and adding terms of the same power provides (optimal) inequalities involving linear combinations of the $t_r$ with $r\le 2n$. A deeper discussion of these difficulties is presented in Section \ref{Sec: Higher spec moments}.

Despite these drawbacks, we find the approach is still viable if we restrict our attention to the first two traces $t_1$ and $t_2$. This is made possible by the existence of two transformations under which the discriminant domain is invariant. They enable a reduction of the integration problem, and open the way to obtain an explicit expression for the JPDF of $(t_1,t_2)$ for the G$\beta$E: \\

 \noindent {\it Spectral shift --}  Shifting every $\lambda_i$ by a constant $\delta\in\mathbb{R}$ changes the traces by
  \begin{equation}\label{eq:traces_shift}
  \tilde t_k=\sum_{i=1}^N(\lambda_i+\delta)^k=\sum_{i=1}^N\sum_{l=0}^k{k\choose l}\delta^l\lambda_i^{k-l}=\sum_{l=0}^k{k\choose l}\delta^lt_{k-l}.
  \end{equation}
  Because the Vandermonde determinant $V(\blambda)$ is invariant under such a shift, we have the invariance of the discriminant
  \[G^2(\tilde\bt)=G^2(\bt).\]
  Recall that the boundary of the integration domain is defined as the zero set of the discriminant, therefore the indicator function is also invariant under such a spectral shift
  \[\chi(\tilde\bt)=\chi(\bt).\]
  (\ref{eq:traces_shift})  shows that the Jacobian of the map $\bt\rightarrow\tilde\bt$ is an upper-triangular matrix with ones on the diagonal, hence its determinant is equal to one. \\

  \noindent {\it Spectral scaling --} Scaling  each of the eigenvalues by the same number $c\in\mathbb{R}$
  scales each trace as
  $t_k'=c^kt_k.$
  The discriminant therefore scales as
  \begin{equation}
  \label{discriminant_scaling}
  G'=\prod_{i<j}(c\lambda_j-c\lambda_i)=c^{N(N-1)/2}\prod_{i<j}(\lambda_j-\lambda_i)=c^{N(N-1)/2}G.
  \end{equation}
  Because the zero set of the discriminant remains unchanged, the indicator function is invariant under such a transformation
  \[\chi(\bt')=\chi(\bt).\]
  The Jacobian of the transformation is  $ c^{N(N+1)/2}$.

\section{The joint probability density function for $t_1,t_2$}\label{JPDF for t1t2}
The above symmetries mean we have two free variables that we can utilise. In particular this allow us to obtain a special coordinate transformation for which $t_1$ and $t_2$ are invariant and the vandermonde (and therefore also the characteristic function $\chi$) are independent of $t_1$ and $t_2$. To see this let us first set $t_1=0$ and $t_2=1$ and then employ our transformations $\delta$ and $c$, as above, to give a new set of coordinates
\begin{eqnarray}\label{Coord change}
\tilde{t}_1' & = & N\delta \nonumber \\
\tilde{t}_2' & = & c^2+N\delta^2, \nonumber \\
\tilde{t}_k' & = & \sum_{l=0}^{k-3}{k\choose l}\delta^lc^{k-l}t_{k-l} + \frac{k(k-1)}{2}\delta^{k-2}c^2 + N\delta^k, \ \  k=3,\ldots N.
\end{eqnarray}
In order to make this coordinate transformation complete we can utilise the freedom afforded to us by the shift $\delta$ and scaling $c$. In other word these variables must depend on the variables $t_1$ and $t_2$.

Crucially, if we choose this dependence in the following form
\begin{equation}
\label{values}
\delta=\frac{t_1}{N}{\rm\ \ and\ \ } c=\sqrt{t_2-\frac{t_1^2}{N}},
\end{equation}
then this gives us mapping $\bt = (t_1,\ldots,t_N) \mapsto \tilde{\bt}' = (\tilde{t_1}',\ldots,\tilde{t}_M')$ defined by (\ref{Coord change}) such that $\tilde{t}_1'= t_1$ and  $\tilde{t}_2'= t_2$. Moreover, These shifts mean that from the discriminant scaling (\ref{discriminant_scaling}) we find
\begin{equation}\label{VDM scaling}
G(\bt)=c^{N(N-1)/2}G(0,1,\tilde t_3',\dots,\tilde t_N'),
\end{equation}
with $c$ given in (\ref{values}). Similarly, because the indicator $\chi$ is given by the zeros of the Vandermonde, we have the relation $\chi(\bt) = \chi(0,1,\tilde t_3',\dots,\tilde t_N')$. 

To calculate the Jacobian of the transformation we note the change of variables (\ref{Coord change}) is given by an lower triangular matrix and therefore multiplying the diagonal elements we find $\prod_{k=3}^N c^{k} = c^{(N^2 + N - 6)/2}$. This allows us to calculate the expectation of a function $f(t_1,t_2)$ and in turn the jpdf of $t_1$ and $t-2$ as follows
\begin{eqnarray}\label{First expectation}
\fl \E[f(t_1,t_2)] & \propto & \int dt_1dt_2\exp\left(-\frac{\beta}{2} t_2\right)f(t_1,t_2)\int dt_3\dots dt_N\chi(\bt)G(\bt)^{\beta-1} \nonumber \\
& = & \int_{t_2\geq t_1^2/N} dt_1dt_2\exp\left(-\frac{\beta}{2} t_2\right)f(t_1,t_2) \mc{C}_{\beta}^{(N)}(t_1,t_2) \nonumber \\
& & \ \ \times  \ \int d\tilde{t}'_3\dots d\tilde{t}'_N\chi(0,1,\tilde{t}_3,\dots,\tilde{t}_N)G(0,1,\tilde{t}_3,\dots,\tilde{t}_N)^{\beta-1}
\end{eqnarray}
where we have used that $t_1 = \tilde{t}'_1$ and $t_2 = \tilde{t}'_2$. Here the factor
\[
\mc{C}_{\beta}^{(N)}(t_1,t_2) = c^{(N^2 + N - 6)/2} c^{(\beta-1)N(N-1)/2} = \left(t_2-\frac{t_1^2}{N}\right)^{\frac{1}{4}(\beta N^2+(2-\beta)N-6)}.
\]
comes from the Jacobian and the scaling (\ref{VDM scaling}). We now notice that the second integral in (\ref{First expectation}) is independent of $(t_1,t_2)$ and can thus be incorporated into the normalisation factor. Doing so we obtain
\[\fl \E[f(t_1,t_2)]\propto\int dt_1dt_2\chi(t_1,t_2)\left(t_2-\frac{t_1^2}{N}\right)^{\frac{1}{4}(\beta N^2+(2-\beta)N-6)}\exp\left(-\frac{\beta}{2} t_2\right)f(t_1,t_2),
\]
where $\chi(t_1,t_2)$ is the indicator function for the domain $\{ (t_1,t_2): t_2 \geq t_1^2/N \}$, coming from (\ref {schwartz_even}). The joint distribution of $t_1$ and $t_2$ is thus given by
\begin{equation}\label{eq:t1t2distribution}
Q^{(N)}_{\beta} (t_1,t_2)=\frac{1}{\mathcal{N}^{(N)}_{\beta} }\chi(t_1,t_2)\left(t_2-\frac{t_1^2}{N}\right)^{\frac{1}{4}(\beta N^2+(2-\beta)N-6)}\exp\left(-\frac{\beta}{2} t_2\right).
\end{equation}
This leaves us to calculate the normalization constant, given by the integral
\[\mathcal{N}^{(N)}_{\beta} =\int_{-\infty}^\infty dt_1\int_{t_1^2/N}^\infty dt_2\left(t_2-\frac{t_1^2}{N}\right)^{\frac{1}{4}(\beta N^2+(2-\beta)N-6)}\exp\left(-\frac{\beta}{2} t_2\right).\]
Setting $u=t_2-t_1^2/N$, we have
\begin{eqnarray*}
\fl \int_{t_1^2/N}^\infty dt_2\left(t_2-\frac{t_1^2}{N}\right)^{\frac{1}{4}(\beta N^2+(2-\beta)N-6)}\exp\left(-\frac{\beta}{2} t_2\right)=\int_{0}^\infty u^{\frac{1}{4}(\beta N^2+(2-\beta)N-6)} \\ \times\exp\left(-\frac{\beta}{2} u\right)\exp\left(-\frac{\beta}{2N} t_1^2\right)du=\Gamma(p+1)\left(\frac{2}{\beta}\right)^{p+1}\exp\left(-\frac{\beta}{2N} t_1^2\right),
\end{eqnarray*}
where $p=\frac{1}{4}(\beta N^2+(2-\beta)N-6)$. The final result is
\begin{equation}\label{normalisation}
\fl \mathcal{N}^{(N)}_{\beta} =\Gamma(p+1)\left(\frac{2}{\beta}\right)^{p+1}\int_{-\infty}^\infty dt_1\exp\left(-\frac{\beta}{2N} t_1^2\right)=\Gamma(p+1)\left(\frac{2}{\beta}\right)^{p+1}\sqrt{\frac{2\pi N}{\beta}}.
\end{equation}

\subsection{Marginal distributions}\label{Marginal distributions}

As an immediate by-product of calculating $\mathcal{N}^{(N)}_{\beta} $ we find that the distribution of $t_1$ is Gaussian
\begin{equation}\label{t1distribution}
Q^{(N)}_{\beta} (t_1)=\sqrt{\frac{\beta}{2\pi N}}\exp\left(-\frac{\beta}{2N} t_1^2\right),\ t_1\in\mathbb{R}.
\end{equation}
In order to compute the distribution of $t_2$, one has to do the integration over $t_1$
\[Q^{(N)}_{\beta} (t_2)=\frac{2}{\mathcal{N}^{(N)}_{\beta} }\exp\left(-\frac{\beta}{2} t_2\right)\int_{0}^{\sqrt{N t_2}} dt_1\left(t_2-\frac{t_1^2}{N}\right)^p.\]
After the substitution $u=t_2-t_1^2/N$, we have
\[\fl \int_{0}^{\sqrt{N t_2}} dt_1\left(t_2-\frac{t_1^2}{N}\right)^p=\frac{\sqrt{N}}{2}\int_0^{t_2}du \frac{u^p}{\sqrt{t_2-u}}=\frac{\sqrt{\pi N}}{2}\frac{\Gamma(p+1)}{\Gamma(p+\frac{3}{2})}t_2^{p+\frac{1}{2}}.\]
Therefore, the distribution of $t_2$ reads
\begin{equation}\label{t2distribution}
Q^{(N)}_{\beta} (t_2)=\left(\frac{\beta}{2}\right)^{p+\frac{3}{2}}\frac{1}{\Gamma(p+\frac{3}{2})}\exp\left(-\frac{\beta}{2} t_2\right)t_2^{p+\frac{1}{2}},\ t_2\in\mathbb{R}_+.
\end{equation}
Thus, $\beta t_2$ obeys a {\it chi-squared} distribution of degree $p+3$.

 Note that changing the variables in (\ref{eq:t1t2distribution}) to $u=\left(t_2-\frac{t_1^2}{N}\right)$ and $v=t_1$, the distribution factorizes to a Gaussian distribution in $v$ and a chi-squared distribution in $u$
\[Q^{(N)}_{\beta} (u,v)=\frac{1}{\mathcal{N}^{(N)}_{\beta} }u^{p}\exp\left(-\frac{\beta}{2} u\right)\exp\left(-\frac{\beta}{2N} v^2\right),\ v\in\mathbb{R},u>0.\]
In the limit of large $N$ the chi-squared distribution also approaches a Gaussian, albeit the $u$ variable is not a linear function of the traces. However, as we will see in Section \ref{sec:correlations}
\[\frac{\E[t_2-t_1^2/N]}{\E[t_2]}\rightarrow1\]
in the limit of large $N$. Therefore, the variable $u$ behaves as $t_2$ in the same limit, so its distribution converges to a Gaussian.  This is in accordance with the known result that linear combinations of the traces (centered around their mean values) with coefficients taken from the corresponding Chebyshev polynomials are independent Gaussians in the limit of large N \cite{Johansson,Schenker-2007}.

\subsection{Correlations $\E[{t_1^mt_2^n}]$}\label{sec:correlations}

The value of $\E[{t_2}]$ in the case of $\beta=1,2,4$-ensembles was calculated for any $N$ in \cite{Ledoux-2009}. Here, we compute the mean value of $t_1^mt_2^n$ for any $\beta$ and therefore reproduce the already known results for $\E[{t_2}]$ in the case of the three ensembles. Let us begin with choosing the most convenient limits of integration
\[\E[{t_1^mt_2^n}]=\frac{1}{\mathcal{N}^{(N)}_{\beta} }\int_{-\infty}^\infty dt_1t_1^m\int_{t_1^2/N}^\infty dt_2t_2^n\left(t_2-\frac{t_1^2}{N}\right)^p\exp\left(-\frac{\beta}{2} t_2\right).\]
After the substitution $u=t_2-t_1^2/N$, we have
\[\fl \E[{t_1^mt_2^n}]=\frac{1}{\mathcal{N}^{(N)}_{\beta} }\int_{-\infty}^\infty dt_1t_1^m\exp\left(-\frac{\beta}{2N} t_1^2\right)\int_0^\infty du\left(u+\frac{t_1^2}{N}\right)^nu^p\exp\left(-\frac{\beta}{2} u\right)\]
\noindent and expanding the term $\left(u+\frac{t_1^2}{N}\right)^n$ we find
\[\fl \E[{t_1^mt_2^n}]=\frac{1}{\mathcal{N}^{(N)}_{\beta} }\sum_{s=0}^n{n\choose s}N^{s-n}\int_{-\infty}^\infty dt_1t_1^{m+2(n-s)}e^{\left(-\frac{\beta}{2N} t_1^2\right)}\int_0^\infty du \ u^{s+p}e^{\left(-\frac{\beta}{2} u\right)}.\]
Note that the integrals factorise under each summand, therefore, taking into account the parity of the function under each integral over $t_1$, we see that the only non-zero result is for $m=2k$. After integrating over $t_2$ are left with
\[\fl \E[{t_1^{2k}t_2^n}]=\frac{1}{\mathcal{N}^{(N)}_{\beta} }\sum_{s=0}^n{n\choose s}\frac{\Gamma(p+s+1)}{N^{n-s}}\left(\frac{2}{\beta}\right)^{p+s+1}\int_{-\infty}^\infty dt_1t_1^{2(k+n-s)}e^{\left(-\frac{\beta}{2N} t_1^2\right)}.\]
Using the formula for the normalisation, integration over $t_1$ yields
\[\E[{t_1^{2k}t_2^n}]=\frac{N^{k}}{\sqrt{\pi}} \left(\frac{2}{\beta}\right)^{k+n}\sum_{s=0}^n{n\choose s}\frac{\Gamma(p+s+1)\Gamma(k+n-s+\frac{1}{2})}{\Gamma(1+p)}.\]
After the summation, we finally have
\begin{equation}\label{meant1t2}
\E[{t_1^{2k}t_2^n}]=\frac{N^{k}}{\sqrt{\pi}} \left(\frac{2}{\beta}\right)^{k+n}\frac{\Gamma(k+\frac{1}{2})\Gamma(k+n+p+\frac{3}{2})}{\Gamma(k+p+\frac{3}{2})}.
\end{equation}
In particular, the mean value of $t_2$ is
\[\E[{t_2}]=\frac{N^2}{2}+\frac{1}{2\beta}(2-\beta)N.\]
This generalises the expressions obtained in \cite{Ledoux-2009} for arbitrary $\beta$ and thus agrees with those in \cite{Dumitriu-2006,Witte-2014} (see also \cite{Olivier,DES07,CMV15}).

\section{Higher spectral moments}\label{Sec: Higher spec moments}

As mentioned previously, there are difficulties in extending the results to higher traces. The general procedure of calculating the integrals is to first find a proper stratification of the integration domain. Such a stratification is done by performing a series of successive cuts of the domain, i.e.
\[\mc{D}=\mc{D}_{N}\supset\mc{D}_{N-1}\supset\dots\supset\mc{D}_{2}\supset\mc{D}_{1},\]
where $\mc{D}_{i}$ is obtained by keeping $t_1,t_2,\dots,t_{N-i}$ fixed. Then, the integration over $t_{N-i+1}$ is restricted to the  projection of $\mc{D}_{i}$ on the $t_{N-i+1}$-axis. These domains are the union of intervals on the $t_{N-i+1}$ axis, which will be denoted by  $\mc{T}^{(N-i+1)}(t_1,t_2,\dots,t_{N-i})$. The integration domain depends parametrically  on the values $(t_1,t_2,\dots,t_{N-i})$. Using the above notation, the integral can be written in the following way
\[
\fl
\int_{\mc{D}}d\bt f(\bt)=\int_{-\infty}^\infty dt_1\int_{\mc{T}^{(2)}(t_1)}dt_2\int_{ \mc{T}^{(3)}(t_1,t_2)}dt_3\dots
\int_{\mc{T}^{N}(t_1,t_2,\dots,t_{N-1})}dt_Nf(\bt),
\]
where we used the fact that the projection of the domain on $t_1$-axis covers the entire real axis. The main obstacle in this procedure is that the domains $\mc {D}_{N-k}$ are not necessarily smooth and thus finding the projections poses a difficult problem.
 
Meymann \cite{Meymann} proposed a systematic method to address the above problem using Lagrange multipliers. For instance, using this technique we can recover the result of the inequality (\ref{schwartz_even}) for $r=1$ (i.e. $t_2 \geq \frac{t_1^2}{N}$), corresponding to the end-points of ${\mc{T}^{(2)}(t_1)}$. Given any value of $t_1 = \sum_{k=1}^N\lambda_k$, the allowed values for $t_2$ are restricted to the interval defined by the extrema of the Lagrangian
\[\mathcal{L}(\blambda;\mu)=\sum_{k=1}^N\lambda_k^2+ \mu
\left(t_1-\sum_{i=1}^N\lambda_i \right).\]
The Lagrange multiplier $\mu$ is chosen such that $t_1=\sum_{k=1}^N\lambda_k $ at the extrema. Clearly, the extrema occur at
\[2\lambda_k-\mu=0 \ \ \ \forall k\ . \]
There exists a single extremum, which is a minimum, occurring at the point $\mu= \frac{2 t_1}{N}$ and hence the minimum value of $t_2$ is $\frac{t_1^2}{N}$ as claimed, since $t_2$ is not restricted from above.

For finding  ${\mc T}^{(k+1)}(t_1,t_2,\dots,t_{k})$ -  the domain of  integration over $t_{k+1}$ using Meymann's method one introduces a Lagrangian of the form 
\begin{eqnarray*}
\mathcal{L}(\blambda;\mu_1,\dots,\mu_{k}) & = & \sum_{i=1}^N\lambda_i^{k+1}+\mu_1\left(t_1-\sum_{i=1}^N\lambda_i\right)+\frac{1}{2}\mu_2\left(t_2-\sum_{i=1}^N\lambda_i^2\right)+ \nonumber \\ 
& & \dots +\frac{1}{k}\mu_{k}\left(t_{k}-\sum_{i=1}^N\lambda_i^{k}\right).
\end{eqnarray*}
The resulting equation for the extrema reads
\begin{equation}\label{lagrange_general}
(k+1)\lambda_j^{k}-\left(\mu_1+\mu_2\lambda_j+\dots+\mu_{k}\lambda_j^{k-1}\right)=0,\ j=1,\dots,N.
\end{equation}
Such a polynomial can have at most $k$ real roots. Assume that we have found the roots. Then, applying equation (\ref{lagrange_general}) for every root yields a set of linear equations for the multipliers $\mu_1,\dots,\mu_k$. Such a set of equations is nonsingular when equation (\ref{lagrange_general}) has exactly $k$ distinct real solutions. In order to associate the multipliers with the constraints, we can express the traces by the roots of equation (\ref{lagrange_general}). Namely, if the roots are $r_1,\dots,r_k$, then
\begin{equation}\label{traces_roots}
t_l=\sum_{i=1}^k p_i r_i^l,\ l=1,\dots,k,
\end{equation}
where $\{p_i\}$ is a collection of integers that sum up to $N$. The solution is very complicated, as one has to find the collection of $p_i$'s that gives extreme values of $t_{k+1}$ and then express $t_{k+1}$ as a function of $t_1,\dots,t_k$ using equations (\ref{lagrange_general}) and (\ref{traces_roots}). Clearly this method becomes more difficult as $k$ increases and seems intractable for large $N$.  

One can, however, use the Meymann method to find bounds on the integration domain ${\mc D}$. Such bounds are very useful  in numerical simulations, especially when the Monte-Carlo method is used for multi-dimensional integrals. To this end, we simplify the problem by, for example, considering only a constraint of $t_2$. Then the Lagrangian reads
\begin{eqnarray*}
\mathcal{L}(\blambda;\mu)=\sum_{i=1}^N\lambda_i^{k+1}+\frac{1}{2}\mu\left(t_2-\sum_{i=1}^N\lambda_i^2\right)
\end{eqnarray*}
and the resulting equations for the critical points are much simpler
\[(k+1)\lambda_j^{k}-\mu\lambda_j=0.\]
There are only two or three roots, depending on the parity of $k$
\[r_1=0,\ r_2=\left(\frac{\mu}{k+1}\right)^{\frac{1}{k-1}},\ r_2'=-\left(\frac{\mu}{k+1}\right)^{\frac{1}{k-1}}.\]
For $t_2 >0$ we have $t_2 = p_2(\mu/(k+1))^{2/(k-1)}$, with $p_2=1,\ldots,N$. Therefore, for $k=2n-1$, we find the extrema of $\mathcal{L}(\blambda,\mu)$ occur at the points $t_{2n} = p_2(\mu/(k+1))^{2n/(k-1)} = p_2^{1-n}t_2^n$. Similarly, for $k=2n$, if we take all the positive roots we find $t_{2n+1} = p_2^{1-n}t_2^{n+1/2}$. Hence, using the allowed values of $p_2$ results in the following inequalities
 \begin{equation}\label{t2_est}
 |t_{2n+1}|\leq (t_2)^{n+\frac{1}{2}},\ \  N^{1-n}(t_2)^{n}\leq t_{2n}\leq (t_2)^{n}.
 \end{equation}
These inequalities can be interpreted as projections of the $t_2$-cut, $\mc{D}_{N-1}(t_2)$, on different axes. Therefore the proper limits of integration are always contained within these bounds. This in turn allows for upper bounds on the expectation values $\E[t_n]$. Unfortunately these are very rough, as the error grows fast with the dimension $N$. For example, one can compare the volume defined by inequalities (\ref{t2_est}) weighted by $e^{-t_2/2}$ with the proper volume of the integration domain, which is given by the normalisation constant for $\beta=1$ in (\ref{traceJPD}). The ratio of the two volumes grows super-exponentially with $N$.

\section{Conclusions}\label{Conclusions}

In summary, we have shown how to compute the JPDF of the first two traces of the G$\beta$E by using a novel method based around the analysis of the domain of allowed values for the vector of traces $\bt$. This has been achieved for arbitrary matrix size $N$ and thus extends some of the previous results by Ledoux \cite{Ledoux-2009}. We have also explained some of the main obstacles which prevent this approach from being used as a tool to compute the expectation values of arbitrary observables that, in general, can depend on the traces of a higher order. The difficulties come from the complicated nature of the integration domain $\mc{D}$ - a volume which is independent of the matrix ensemble. We believe the investigation of $\mc{D}$ may pose an interesting problem in Algebraic Geometry and other related domains in mathematics. 

\ack
We would like to acknowledge S. Sodin for bringing to our attention the reference of N. Meymann and other literature. We also thank S. Yakovenko for instructive discussions. TM would like to thank the Weizmann Institute of Science for a great hospitality during his stay in the summer of 2015, when this work was initiated. TM is supported by Polish Ministry of Science and Higher Education ``Diamentowy Grant'' no. DI2013 016543 and European Research Council grant QOLAPS. CHJ would like to thank the Leverhulme Trust (ECF-2014-448) for financial support.


\section*{References}
\begin{thebibliography}{9}


\bibitem{Anderson-book} G. W. Anderson, A. Guionnet, and O. Zeitouni, \emph{An Introduction to Random Matrices}, Cambridge Studies in Advanced Mathematics {\bf 118} (Cambridge University Press, 2009).

\bibitem{Wigner-1955}
E. P. Wigner, \emph{Characteristic vectors of bordered matrices with infinite dimensions}, Ann. Math., {\bf 62}, 548-564 (1955).

\bibitem{Wigner-1958}
E. P. Wigner, \emph{On the Distribution of the Roots of Certain Symmetric Matrices}, Ann. of Math. {\bf 67}, 325-328 (1958). 

\bibitem {Mezzadri1} F. Mezzadri and N. J. Simm, \emph{Moments of the transmission eigenvalues, proper delay times, and random matrix theory. I and II}. J. Math. Phys. {\bf 52}, 103511 (2011) and {\bf 53} (2012), 053504

\bibitem {Mezzadri2}F. Mezzadri and N. J. Simm, \emph{Tau-Function Theory of Chaotic Quantum Transport with $\beta$ = 1, 2, 4}, Commun. Math. Phys. {\bf 324}, 465-513 (2013)

\bibitem{mehtabook}
M. L. Mehta, \emph{Random Matrices}, Third Edition, 142, Pure and Applied Mathematics (Elsevier/Academic Press, Amsterdam, 2004).

\bibitem{Dumitriu}
I. Dumitriu and A. Edelman, \emph{Matrix models for beta ensembles}, J. Math. Phys. {\bf 43}, 5830 (2002) 

\bibitem{Johansson}
K. Johansson, \emph{On fluctuations of eigenvalues of random Hermitian matrices}, Duke Math. J. {\bf 91}, 151-204 (1998).

\bibitem{Dumitriu-2006}
I. Dumitriu and A. Edelman, \emph{Global spectrum fluctuations for the $\beta$-Hermite and $\beta$-Laguerre ensembles via matrix models}, J. Math. Phys. {\bf 47}, 063302 (2006)

\bibitem{Cabanal-Duvillard-2001}
T. Cabanal-Duvillard and A. Guionnet, \emph{Large Deviations Upper Bounds for the Laws of Matrix-Valued Processes and Non-Communicative Entropies}, Ann. Prob. {\bf 29}, No. 3 (Jul., 2001), pp. 1205-1261.

\bibitem{Sinai}
Y. Sinai and A. Soshnikov, \emph{Central limit theorem for traces of large random matrices with independent matrix elements}, Bol. Soc. Bras. {\bf 29} 1-24 (1998).

\bibitem{Khorunzhy-1996}
A. M. Khorunzhy, B. A. Khoruzhenko, L. A. Pastur, \emph{Asymptotic properties of large random matrices with independent entries}, J. Math. Phys. 37, 5033-5060 (1996).

\bibitem{Bai-2005}
Z. D. Bai and J.-F. Yao, \emph{On the convergence of the spectral empirical process of Wigner matrices}, Bernoulli {\bf 11}, 1059-1092 (2005).

\bibitem{Schenker-2007}
J. Schenker and H. Schulz-Baldes, \emph{Gaussian fluctuations for random matrices with correlated entries}, Int. Math. Research Notices {\bf 15}, article ID rnm047, 36 pages (2007).

\bibitem{Diaconis-1994}
P. Diaconis and M. Shahshahani, \emph{On the eigenvalues of random matrices}, J. Appl. Probab. {\bf 31A} (1994), 49-62.

\bibitem{Haake-1996}
F. Haake, M. Ku\'{s}, H.-J. Sommers, H. Schomerus and K. \.{Z}yczkowski, \emph{Secular determinants of random unitary matrices}, J. Phys. A: Math. Gen. {\bf 29} (1996) 3641–3658.

\bibitem{Webb-2015}
C. Webb, \emph{Linear statistics of the circular $\beta$-ensemble, Stein's method and circular Dyson Brownian motion}, Preprint (2015), http://arxiv.org/abs/1507.08670

\bibitem{Ledoux-2009} M. Ledoux, \emph{A recursion formula for the moments of the gaussian orthogonal ensemble}, Annales de l'Institut Henri Poincar\'{e} - Probabilit\'{e}s et Statistiques, Vol. 45, No. 3, 754-769 (2009)

\bibitem{Harer-1986}
J. Harer and D. Zagier, \emph{The Euler characteristic of the moduli space of curves}, Invent. Math. 85 (1986) 457-485. MR0848681

\bibitem{Witte-2014}
N. S. Witte and P. J. Forrester, \emph{Moments of the Gaussian $\beta$-ensembles and the large-$N$ expansion of the densities}, J. Math. Phys. {\bf 55}, 083302 (2014)

\bibitem{Olivier}
Marchal, O., {\it Elements of proof for conjectures of Witte and Forrester about the combinatorial structure of Gaussian $\beta$-Ensembles}, Journal of High Energy Physics, vol. 3 (2014)

\bibitem{DES07}
Dumitriu, I., Edelman, A., Shuman, G., {\it MOPS: Multivariate orthogonal polynomials (symbolically)}, Journal of Symbolic Computation 42 pp. 587?620 (2007)

\bibitem{CMV15}
Cunden, F., Mezzadri, F., Pierpaolo, V., {\it A unified fluctuation formula for one-cut $\beta$-ensembles of random matrices}, Journal of Physics A: Mathematical and Theoretical, Volume 48, Number 31 (2015)

\bibitem{Joyner-2015}
C. H. Joyner and U. Smilansky, \emph{Dyson's Brownian-motion model for random matrix theory - revisited}. With an Appendix by Don Zagier, preprint (2015), http://arxiv.org/abs/1503.06417

\bibitem{Bakry-2014}
D. Bakry and M. Zani, \emph{Dyson processes associated with associative algebras: The Clifford case}, in \emph{Geometric Aspects of Functional Analysis} (eds. B. Klartag, E. Milman), Lecture Notes in Mathematics Volume 2116, pp 1-37 (Springer International Publishing Switzerland 2014).

\bibitem{Dunne-1993}
G. V. Dunne, \emph{Slater decomposition of Laughlin states}, Int. Journ. Mod. Phys. N 7 (28), 4783 (1993).

\bibitem{Vivo-2008}
P. Vivo and S. N. Majumdar, \emph{On invariant $2 \times 2$ $\beta$-ensembles of random matrices}, Phys. A. 387 (2008), no. 19-20, 4839-4855.

\bibitem{Meymann} N. Meymann, \emph{Sur les polynomes R-prolongeables}, Rec. Math. N.S., 1938, Volume 3(45), Number 3, 591-650


%


%

%

%
%
%
%
%
%
%
%
%
%
%
%
\end{thebibliography}
\end{document}